\documentstyle[11pt]{article}

\def\no{\nonumber\\}

\def\roughly#1{\mathrel{\raise.3ex\hbox{$#1$\kern-.75em%
\lower1ex\hbox{$\sim$}}}}

\def\lsim{\roughly<}
\def\gsim{\roughly>}
\def\be{\begin{eqnarray}}
\def\ee{\end{eqnarray}}
\def\susc{susceptibility}
\def\Tr{{\rm Tr}}

\def\ben{\begin{enumerate}}
\def\een{\end{enumerate}}
\def\beitem{\begin{itemize}}
\def\eitem{\end{itemize}}

\newcommand{\beq}{\begin{eqnarray}}
\newcommand{\eeq}{\end{eqnarray}}
\def\la{\langle}
\def\ra{\rangle}
\def\bi{\begin{itemize}}
\def\ei{\end{itemize}}
\def\ie{{\it i.e}}

\def\etal{{\it et al}}
\def\del{\partial}
\def\L{{\cal L}}

\long\def\beginomit#1\endomit{}
\def\np{{Nucl. Phys.}}
\def\prl{Phys. Rev. Lett.}
\def\pr {Phys. Rev.}
\def\PR {Phys. Repts.}

\def\pl{Phys. Lett.}
\def\L{{\cal L}}

\begin{document}


\begin{center}

\vskip 0.4in
{\Large\bf CHIRAL SYMMETRY RESTORATION}
\vskip 0.1cm
{\Large\bf AS THE GEORGI VECTOR LIMIT}\footnote{Invited talk given at
{\it Hirschegg '95: Hadrons in Nuclear Matter}, Hirschegg, Kleinwalsertal,
Austria, January 16-21, 1995.}
\vskip 0.4in
{\large  Mannque Rho}\\
\vskip 0.1in
{\large \it Service de Physique Th\'{e}orique, CEA  Saclay}\\
{\large\it 91191 Gif-sur-Yvette Cedex, France}\\
\vskip .4in
\centerline{January 1995}
\vskip .4in

{\bf ABSTRACT}\\ \vskip 0.6cm
\begin{quotation}

\noindent  I discuss recent work done with Gerry Brown
on chiral phase transition at high temperature and/or density
described in terms of Georgi's vector limit.
The notion of ``mended symmetry" is suggested to play an important
role in understanding the properties of hadrons in dense and/or hot
matter before reaching the phase transition. It is shown that while the
QCD vacuum in baryon-free space is resistant to ``melting" up to the
critical temperature, baryon-rich medium can induce the vacuum to
become softer in temperature: the hadron masses drop faster in
temperature when baryon matter is present.
\end{quotation}
\end{center}

\vskip 0.8cm

\subsubsection*{Introduction}
\indent

At low energies, various hadrons enter into certain physical processes
without apparent relations between them. Thus for instance, the pion-nucleon
scattering near threshold can be described with pions and nucleons alone
or with pions, nucleons and the scalar meson $\sigma$ with a large mass
or pions, nucleons and the vector meson $\rho$. The common element is the
spontaneously broken chiral symmetry which puts constraints so as to make
all these diverse approaches give the same answer. At low energies,
there is no obvious
raison-d'\^etre for the presence of the $\sigma$ or the $\rho$.
In this talk, I would like to argue that as one increases temperature
and/or density of the hadronic system as in heavy-ion collisions or
compact stars, the $\pi$, the $\sigma$, the $\rho$ etc. come together
into a pattern that reveals ``hidden symmetries."  The reasoning is
very much like what Weinberg calls ``mended symmetry"\cite{weinbergMS}
and involves in particular the notion of the ``vector limit" proposed by
Georgi\cite{georgi} some years ago. Both Weinberg and Georgi
invoke the large $N_c$ limit
as the appropriate limit. In papers recently published by us\cite{br94}
and to appear\cite{newbr}, we proposed
that high T and/or high density would bring the matter
to the corresponding limits without involving explicit large $N_c$ assumption.

To illustrate the main point, I discuss three cases: 1) the role of the
$\sigma$ meson in nuclear physics;
2) the quark-number susceptibility at high temperature measured in
lattice gauge calculations; 3) ``cool" kaons observed in
heavy-ion collisions.
\subsubsection*{Mended symmetry and the linear $\sigma$ model}
\indent

At zero temperature ($T=0$) and zero density ($\rho=0$), low-energy
pion interactions are described by the nonlinear $\sigma$ model.
No low-mass scalars ({\ie}, the $\sigma$) figure in the description
and pions are derivatively coupled. There is of course a scalar associated
with the trace anomaly of QCD
\be
\chi\sim (G^2_{\mu\nu})^{1/4}
\ee
where $G_{\mu\nu}$ is the gluon field tensor, but this involves
a high excitation with its mass greater than the chiral scale
$\Lambda_\chi\sim 1$ GeV and so decouples from the low-energy
dynamics. Thus for instance, pion-nucleon scattering at low energy
can be described by a Lagrangian that contains the nucleon and the
pion field but no scalar particle.

However when one studies nucleon interactions in nuclei, one finds that
there is a definite need for a scalar particle $\sigma$ that gives the
principal attraction between nucleons. More precisely this may be just
a scalar field that interpolates $2\pi$, $4\pi$ etc. correlations.
This can be understood as follows.

Suppose that nuclear environments provide kinematic situation where
higher energy (or momentum) scale is involved. This suggests imposing
certain high energy constraints on the low-energy properties embodied in the
current algebra Lagrangian of the nonlinear $\sigma$ model. Weinberg
has considered the constraints provided by Regge asymptotic properties
and showed that certain symmetries are ``mended" although the system
is still in a broken phase. He thus identifies the quartet
$\pi$, $\epsilon$, $\rho_0$ and $a_{1_0}$ (with the subscript
0 denoting the helicity-0 component) to satisfy a mended symmetry
relation. Using this argument, Beane and van Kolck\cite{beane}
suggested that such high-energy constraints bring the glueball field $\chi$
down in  energy and turn it into a dilaton $\sigma$ with its mass nearly
degenerate with the pion mass\footnote{To be more precise, it is the
quarkish component of the $\chi$ field that becomes a dilaton as clarified
in \cite{newbr}. The non-smooth (high-frequency)
glueball component is integrated out and
does not figure explicitly in low-energy dynamics.}.
The symmetry mended in this way
-- while the mass of the scalar is not quite degenerate with the pion
-- is then the $O(4)$ symmetry involving the quartet $\pi^1$,
$\pi^2$, $\pi^3$ and $\sigma$. The resulting Lagrangian is precisely the
linear $\sigma$ model, apart from a logarithmic potential term that
arises from the incorporation of the QCD trace anomaly. This Lagrangian
is then assumed to be applicable at some density not far from a
critical point. We suggest that this model is applicable
already at nuclear matter density with the scalar mass $m_\sigma\sim
400$ MeV, the meson that figures in Walecka's model and also in
one-boson-exchange potentials.

I should mention that this work gives support to the scaling proposed
by us\cite{br91} for the hadrons in medium.

Assuming that the Beane-van Kolck model describes a dense medium
with its constants approaching the dilaton limit, an interesting question
is how hadrons respond to temperature. We know from chiral perturbation
theory (supported by lattice calculations) that the properties of
hadrons in zero-density space
do not change appreciably up to the chiral phase transition temperature
$T_c\approx 140$ MeV. For instance, at low temperature, chiral perturbation
theory (or the linear $\sigma$ model) predicts that the pion decay
constant has the temperature dependence
\be
\frac{f_{\pi} (T)}{f_{\pi}} =1-\frac{T^2}{12f_{\pi}^2}+\cdots\label{fpit}
\ee
Now consider the Beane-van Kolck
Lagrangian in the dilaton limit given by
\be
L &=& i\bar Q \not\! \partial Q
+ \frac{1}{2} \partial_\mu \vec{\pi} \partial^\mu \vec{\pi}+
\frac{1}{2}\partial_{\mu}\sigma\partial^{\mu}\sigma
-\frac{m^\star}{f_{\pi}}\bar Q [\sigma -i\gamma_{5}\vec \pi
\cdot\vec\tau ]Q \no
 &+& \frac{{m^\star_{\sigma}}^2}{16 {f^\star_{\pi}}^2}(\sigma^2 + \vec\pi^2)^2
-\frac{{m^\star_{\sigma}}^2}{8 {f^\star_{\pi}}^2}[(\sigma^2 + \vec\pi^2)^2
\ln(\sigma^2 + \vec\pi^2) /{f^\star_{\pi}}^2]\label{dchmodel}
\ee
where $Q$ is the constituent quark field with mass $m^\star$, $\sigma$
the dilaton field with mass $m_\sigma^\star$ and the pion decay constant
$f_\pi^\star$ is the in-medium constant given by the Brown-Rho scaling.
With this Lagrangian, the temperature dependence of the pion decay constant
is calculated to be\cite{KLR}
\be
\frac{f^\star_{\pi} (T)}{f^\star_{\pi}}
=1-\frac{T^2}{4{f^\star_{\pi}}^2}+\cdots\label{fpitstar}
\ee
This shows that in the presence of baryonic matter, the pion decay constant
drops substantially
faster in temperature than in baryon-free space (\ref{fpit}). (Note that
at $\rho\approx \rho_0$, $f_\pi^\star/f_\pi\approx 0.8.$)
One can also show\cite{KLR} that the BR scaling holds approximately
in this model as a function of temperature. It would be interesting to
check this property in lattice gauge calculations.

\subsubsection*{The Georgi vector limit}
\indent

In a way analogous to the appearance of the low-mass scalar $\sigma$
discussed above,
vector mesons can also be brought into the low-energy sector. This involves
the Georgi vector symmetry
and vector limit\cite{georgi}. Consider two chiral flavors u(p) and d(own),
with chiral symmetry $SU(2)_L\times SU(2)_R$. The standard way of looking at
this symmetry is that it is realized either in Nambu-Goldstone (or Goldstone
in short) mode, with $SU(2)_L\times SU(2)_R$ broken down spontaneously to
$SU(2)_{L+R}$ or in Wigner-Weyl (or Wigner in short) mode with parity
doubling. Georgi observes, however, that
there is yet another way of realizing the symmetry which requires
both Goldstone mode and Wigner mode to {\it co-exist}. Now the signature
for {\it any} manifestation of the chiral symmetry is the pion decay constant
$f_\pi$
\be
\la 0 |A^i_\mu|\pi^j (q)\ra=i f_\pi q_\mu \delta^{ij}\label{goldstone}
\ee
where $A_\mu^i$ is the isovector axial current. The Goldstone mode is
characterized by the presence of the triplet of Goldstone bosons,
$\pi^i$ with $i=1, 2, 3$ with a non-zero pion decay constant.
The Wigner mode is realized when the pion decay constant vanishes, associated
with
the absence of zero-mass bosons. In the latter case, the symmetry is realized
in a parity-doubled mode. The Georgi vector symmetry we are interested in
corresponds to
the mode (\ref{goldstone}) co-existing with a triplet of scalars $S^i$
with $f_S=f_\pi$ where
\be
\la 0|V_\mu^i| S^j (q)\ra= i f_S q_\mu \delta^{ij}
\ee
where $V_\mu^i$ is the isovector-vector current. In this case, the
$SU(2)\times SU(2)$ symmetry is unbroken. At low $T$ and/or low density,
low-lying isovector-scalars are {\it not} visible and hence either the vector
symmetry is broken in Nature with $f_S\neq f_\pi$ or they are ``hidden"
in the sense that they are eaten up by vector particles (\`a la Higgs).
We are suggesting that as temperature and/or density
rises to the critical value corresponding to the chiral phase transition,
the symmetry characterized by
\be
f_S=f_\pi\label{equal}
\ee
is restored with the isovector scalars making up the longitudinal components
of the massive $\rho$ mesons, which eventually get ``liberated" at some high
temperature (or density) from the
vectors and become degenerate with the zero-mass
pions at $T\gsim T_{\chi SR}$ where
$T_{\chi SR}\sim T_c\sim 140$ MeV is the chiral transition temperature.
The symmetry (\ref{equal}) with the scalars ``hidden" in the massive vector
mesons resembles Weinberg's mended symmetry\cite{weinbergMS}.
We shall reserve
``Georgi vector limit" as the symmetry limit in which (\ref{equal}) holds
together with $m_\pi=m_S=0$.

The relevant Lagrangian to use for illustrating
what we mean is the hidden gauge
symmetric Lagrangian of Bando {\etal}\ \cite{bando} which is valid below
the chiral transition\footnote{We will ignore the small quark masses and
work in the chiral limit.},
\be
\L=\frac 12 f^2\left\{\Tr(D^\mu\xi_L D_\mu \xi_L^\dagger) +
(L\rightarrow R)\right\} +\kappa \cdot \frac 14 f^2 \Tr (\del^\mu U\del_\mu
U^\dagger) +\cdots\label{bandoL}
\ee
where $U=\xi_L \xi_R^\dagger$, $D^\mu \xi_{L,R}
=\del^\mu\xi_{L,R}-ig\xi_{L,R}\rho^\mu$, $\rho_\mu\equiv
\frac 12 \tau^a \rho^a_\mu$
and $g$ stands for the hidden gauge coupling. The ellipsis stands for other
matter fields and higher-derivative terms needed to make the theory more
realistic.
The $\xi$ field can be parametrized as
$\xi_{L,R}\equiv e^{iS(x)/f_S} e^{\pm i\pi (x)/f_\pi}$
with $S(x)=\frac 12 \tau^a S^a (x)$ and $\pi (x)=\frac 12 \tau^a \pi^a (x)$.
The symmetry of the Lagrangian (\ref{bandoL}) is
$(SU(2)_L\times SU(2)_R)_{global} \times G_{local}$ with $G\in SU(2)_V$.
Setting $S(x)=0$ corresponds to taking the unitary gauge in which case
we are left with physical fields only ({\ie}, no ghost fields).

At tree level, we get that
\be
f_S=f, \ \ f_\pi=\sqrt{1+\kappa} f
\ee
and the $\rho\pi\pi$ coupling
\be
g_{\rho\pi\pi}=\frac{1}{2(1+\kappa)} g.
\ee
Going to the unitary gauge, one gets the KSRF mass relation
\be
m_\rho=fg=\frac{1}{\sqrt{1+\kappa}} f_\pi g =
2\sqrt{1+\kappa} f_\pi g_{\rho\pi\pi}.\label{KSRF}
\ee
We know from experiments that at zero $T$ (or low density), the
$\kappa$ takes the value $-\frac 12$ for which the KSRF relation is
accurately satisfied. The symmetry (\ref{equal}) therefore is broken.
The symmetry is recovered
for $\kappa=0$ in which case the second term of (\ref{bandoL})
that mixes L and R vanishes, thus restoring $SU(2)\times SU(2)$. In this
limit, the hidden gauge symmetry swells to $G_L\times G_R$, and $\xi_{L,R}$
transform $
\xi_L\rightarrow L\xi_L G_L^\dagger, \ \ \xi_R\rightarrow R\xi_R G_R^\dagger.$
If the gauge coupling is not zero, then the $\rho$ mesons are still
massive and we have the mended symmetry (\ref{equal}). However if the
gauge coupling vanishes, then the vector mesons become massless and
their longitudinal components get liberated, giving the scalar massless
multiplets $S(x)$. In this limit, the symmetry is the global $[SU(2)]^4$.
Local symmetry is no longer present.

It is our proposal that in hot and/dense matter approaching the chiral
restoration, the constant $\kappa\rightarrow 0$ and the gauge coupling
$g\rightarrow 0$.  For this purpose, we shall extrapolate a bit
the results obtained by Harada and Yamawaki\cite{harada}. These authors
studied the hidden gauge Lagrangian (\ref{bandoL}) to one loop order
and obtained the $\beta$ functions for the hidden gauge coupling $g$
and the constant $\kappa$ (in dimensional regularization)
\be
\beta_g (g)&=& \mu \frac{dg}{d\mu}=-\frac{87-a^2}{12} \frac{g^2}{(4\pi)^2},
\\
\beta_\kappa (a)&=& \mu \frac{da}{d\mu}= 3a (a^2-1) \frac{g^2}{(4\pi)^2}
\ee
with $a=\frac{1}{1+\kappa}$. One notes that first of all, there
is a nontrivial ultraviolet fixed point at $a=1$ or $\kappa=0$
and that the coupling constant $g$ scales to zero as $\sim 1/\ln \mu$ in the
ultraviolet limit. This perturbative result may not be realistic
enough to be taken seriously -- and certainly cannot be pushed too high in
energy-momentum scale but for the reason given below, we think it
plausible that the Harada-Yamawaki results hold at least qualitatively as
high $T$ (or density) is reached.
In fact we expect that the gauge coupling should fall off to zero much faster
than logarithmically in order to explain what follows below.

\subsubsection*{Quark-number \susc}
\indent

As a case for the role of the vector limit, we consider the
lattice gauge calculations made by Gottlieb {\etal}
\cite{gottliebchi} of the quark-number susceptibility defined by
\be
\chi_{\pm}=\left(\del/\del \mu_{u} \pm \del/\del \mu_d\right) (\rho_u\pm
\rho_d)
\ee
where the $+$ and $-$ signs define the singlet (isospin zero) and triplet
(isospin one) susceptibilities, $\mu_u$ and $\mu_d$ are the chemical
potentials of the up and down quarks and
\be
\rho_i=\Tr N_i exp\left[-\beta (H-\sum_{j=u,d} \mu_j N_j)\right]/V
\equiv \la\la N_i\ra\ra/V
\ee
with $N_i$ the quark number operator for flavor $i=u,d$.
The $\chi_+$ is in the $\omega$-meson channel and
the $\chi_-$ in the $\rho$-meson channel.
Due to the $SU(2)$ symmetry, $\chi_+=\chi_-$ as one can see
in the lattice results.
Since the singlet susceptibility is similar
to the non-singlet one, we consider the latter.

One can classify the lattice results by roughly
three temperature regimes.
In the very low temperature regime, the $\chi_-$ is dominated
by the $\rho$ meson and is small. As the temperature moves toward the onset of
the phase transition, the $\chi_-$ increases rapidly to near that of
non-interacting quarks. This regime may be described in terms of constituent
quarks. In RPA approximation of the constituent quark model as used
by Kunihiro\cite{kunihiro}, the susceptibility below the critical
temperature is
\be
\chi=\frac{\chi_0}{1+G_v \chi_0}
\ee
where $G_v$ is the coupling of the constituent quark (denoted $Q$) to the
vector meson $\rho$ and $\chi_0$ is the susceptibility for non-interacting
quarks which at $T\approx T_{\chi SR}$ where the dynamical mass $m_Q$ has
dropped to zero has the value
\be
\chi_0\approx N_f T^2
\ee
with $N_f$ the number of flavors. In terms of the gauge coupling of
(\ref{bandoL}), we have
\be
G_v\approx \frac{g^2}{4m_\rho^2}.
\ee
As noted by Kunihiro in the NJL model, the rapid increase of the
\susc\  can be understood by a steep drop in the vector coupling across
the $T_{\chi SR}$. Let us see what we obtain with the hidden gauge
symmetry Lagrangian (\ref{bandoL}). If we assume that the KSRF
relation (\ref{KSRF}) holds at $T$ near $T_{\chi SR}\approx 140$ MeV
(the recent work by Harada {\etal}\ \cite{haradaPRL} supports this
assumption) and that $\chi_0\approx 2T_{\chi SR}^2$ for $N_f=2$, then we find
\be
\chi (T_{\chi SR})/\chi_0 (T_{\chi SR})\approx \frac{1}{1+\frac 12
(\frac{T_{\chi SR}}{f_\pi})^2} \approx 0.47\label{HGSchi}
\ee
with $\kappa=0$.
Here we are assuming that $f_\pi$ remains at its zero temperature value,
93 MeV, up to near $T_{\chi SR}$. The ratio (\ref{HGSchi})
is in agreement with the lattice data at $T\lsim T_{\chi SR}$.

Let us finally turn to the third regime, namely above
$T_{\chi SR}$.
It has been shown by Prakash and Zahed \cite{prakash} that with increasing
temperature, the susceptibility goes to its perturbative value which can
be calculated with perturbative gluon-exchanges.
The argument is made with the dimensional reduction at asymptotic temperatures,
but it seems to apply even at a temperature slightly above $T_{\chi SR}$.
We shall schematize
the Prakash-Zahed argument using the dimensionally reduced model of Koch
{\etal}\ \cite{KSBJ} which hints at the onset of the Georgi vector limit.
In this model which exploits the ``funny space" obtained by interchanging
$z$ and $t$, the helicity-zero state of
the $\rho$ meson is found to come out degenerate with the pion while the
helicity $\pm$ states are degenerate with each other. In finite temperature,
$z$ replaces $T$, so asymptotically in $T$, the configuration space with the
new $z$ becomes 2-dimensional with $x$ and $y$. The $\rho$ meson has gone
massless and behaves like a (charged) photon with helicities $\pm$ 1
perpendicular to the plane. The helicity-zero state originating from
the longitudinally polarized component of the $\rho$ before it went massless
now behaves as an isotriplet scalar. We identify this with the scalar $S(x)$
described above, a realization of the Georgi vector symmetry.

Let us assume then that the vector mesons have decoupled with $g=0$. Going to
the perturbative picture with quark-gluon exchanges, we take
one-gluon-exchange potential of Koch {\etal},
\be
V(r_t)=\frac{\pi}{m^2}\frac 43 \bar{g}^2 T
\sigma_{z,1}\sigma_{z,2}\delta (r_t)
\label{V}
\ee
with $\bar{g}$ the color gauge coupling and $\delta (r_t)$ is the
$\delta$-function in the two-dimensional reduced space. Here $m=\pi T$ is
the chiral mass of quark or antiquark as explained in \cite{KSBJ}.
Possible constant terms that can contribute to eq.(\ref{V}) will be ignored
as in \cite{KSBJ}.
In order to evaluate the expectation value of the $\delta (r_t)$, we note that
the helicity-zero
$\rho$-meson wave function in two dimensions is well approximated by
$\psi_\rho\approx N e^{-r_t/a}$
with $a\approx \frac 23$ fm and the normalization
$N^2=\frac{2}{\pi a^2}.$
For the helicity $\pm 1$ $\rho$-mesons, $\sigma_{z,1}\sigma_{z,2}=1$,
so we find that the expectation value of $V$ is
\be
\langle V\rangle=\frac 83 \frac{\bar{g}^2 T}{\pi^2 T^2 a^2}.
\ee

Now summing the ladder diagrams to all orders, we get
\be
\frac{\chi}{\chi_0}=\left(1+\frac{\langle V \rangle}{2\pi T}\right)^{-1},
\label{ratio}
\ee
where the energy denominator $2\pi T$ corresponds to the mass of a pair
of quarks.

The lattice calculations \cite{gottliebchi} use $6/\bar{g}^2=5.32$
which would give $\alpha_s=0.07$ at scale of $a^{-1}$ where $a$ is the lattice
spacing. Calculations
use  4 time slices, so the renormalized $\bar{g}$ is that appropriate
to $a^{-1/4}$. Very roughly we take this into account by multiplying the
above $\alpha_s$ by $\ln 4^2$; therefore using $\alpha_s\cong 0.19$.
With this $\alpha_s$ and the above wave function, we find
\be
\frac{\chi (T_{\chi SR}^+)}{\chi_0 (T_{\chi SR}^+)}\approx 0.68.
\label{pertchi}
\ee
This is just about the ratio obtained above $T_{\chi SR}$
in the lattice calculations. Remarkably the perturbative result
(\ref{pertchi}) above $T_c$ matches smoothly onto
the HGS prediction (\ref{HGSchi}) just below $T_c$. Neglecting logarithmic
dependence of the gauge coupling constant, eq. (\ref{ratio}) can be
written as
\be
\frac{\chi}{\chi_0} (T)\approx \frac{1}{1+0.46 (T_c/T)^2}
\ee
which follows closely the lattice gauge results of
Gottlieb {\etal}\ \cite{gottliebchi}. We consider this an
indication for the Georgi vector symmetry, with the
induced flavor gauge symmetry in the hadronic sector ceding to the
fundamental color gauge symmetry of QCD in the quark-gluon sector.

We should remark that to the extent that the screening mass obtained in
\cite{KSBJ} $m_\pi=m_S\approx 2\pi T$ is consistent with two non-interacting
quarks and that the corresponding wave functions obtained therein are
the same for the pion $\pi$ and the scalar $S$, we naturally expect the
relation (\ref{equal}) to hold.

\subsubsection*{Cool kaons in heavy-ion collisions}
\indent

The vanishing of the hidden gauge coupling $g$ can have a dramatic effect
on the kaons produced in relativistic heavy-ion collisions. In particular,
it is predicted that the kaons produced from quark-gluon plasma would have a
component that has
a temperature much lower than that for other hadrons. This
scenario may provide an explanation of the recent preliminary
data\cite{stachel} on the $14.6$ GeV collision (experiment E-814)
\be
^{28}{\rm Si} + {\rm Pb}\rightarrow K^+ (K^-) + X
\ee
which showed cool components with effective
temperature of 12 MeV for $K^+$ and 10 MeV for $K^-$, which cannot be
reproduced in the conventional scenarios employed in event generators.
The latter give kaons of effective temperature $\sim 150$ MeV.

There are two points to keep in mind in understanding what is happening
here. Firstly, the Brookhaven AGS experiments determined the freeze-out --
the effective decoupling in the sense of energy exchange of pions and
nucleons -- at $T_{fo}\approx 140$ MeV\cite{BSW}.
This is essentially the same as
the chiral transition temperature measured in lattice gauge calculations
\cite{lattice}. This suggests that the freeze-out for less strongly interacting
particles other than
the pion and the nucleon is at a temperature higher than $T_{\chi SR}$ and
that the pion and nucleon freeze out at about $T_{\chi SR}$. This means
that interactions in the interior of the fireball will be at temperature
greater than $T_{\chi SR}$. At this temperature, the vector coupling
$g$ would have gone to zero, so the Georgi vector limit would be
operative were it to be relevant.
The second point is that the fireball must expand slowly.
The slow expansion results because the pressure in the region
for some distance above $T_{\chi SR}$ is very low \cite{kochbrown}, the
energy in the system going into decondensing gluons rather than giving
pressure.
This results in an expansion velocity of $v/c\sim 0.1$.
In the case of 15 GeV/N Si on Pb transitions, the fireball has been measured
\cite{braun} through Hanbury-Brown-Twiss correlations of the pions to increase
from a transverse size of $R_T (Si)=2.5$ fm to $R_T=6.7$ fm, nearly a factor
of 3, before pions freeze out. With an expansion velocity of $v/c\sim 0.1$,
this means an expansion time of $\sim 25 - 30$ fm/c. (The full expansion
time cannot be measured from the pions which occur as a short flash at the
end.)

In a recent paper, V. Koch\cite{koch}
has shown that given a sizable effective attractive
interaction between the $K^+$ and the nucleon at the freeze-out phase,
a cool kaon component can be reproduced in the conditions specified above.
He elaborated further
on this in this meeting. We argue now that such an attractive
interaction can result if the Georgi vector limit is realized.

The description by chiral perturbation theory\cite{knpw,LJMR,LBR,LBMR}
of kaon nuclear interactions and kaon condensation in
dense nuclear matter has shown that three mechanisms figure prominently
in kaon-nuclear processes at low energy: (1) the $\omega$ meson exchange
giving rise to repulsion for $K^+ N$ interactions and attraction
for $K^- N$; (2) the ``sigma-term" attraction for both $K^\pm N$:
(3) the repulsive ``virtual pair term." In effective chiral
Lagrangians, the first takes the form, $\sim \pm
\frac{1}{f^2} K^\dagger \del_\mu K \bar{N} \gamma^\mu N$ for $K^\pm$,
the second $\sim \frac{\Sigma_{KN}}{f^2} K^\dagger K \bar{N} N$
and the third term $\sim (\del_\mu K)^2 \bar{N}N$.
The vector-exchange gives the repulsive
potential
\be
V_{K^+ N}\cong \frac 13 V_{NN}\cong 90\ {\rm MeV}\,\frac{\rho}{\rho_0}
\label{repulsion}
\ee
where $\rho_0$ is nuclear matter density. It is not explicit but
this term is proportional to
the hidden gauge coupling $g^2$. As for the scalar attraction, it is mainly
given by the ``sigma term"
\be
S_{K^+ N}\approx -\frac{\Sigma_{KN} \langle \bar{N}N\rangle}
{2 m_K f^2}\cong -45\ {\rm MeV}\,\frac{\rho_s}{\rho_0}
\label{attraction}
\ee
where $\rho_s$ is the scalar density and $\Sigma_{KN}$ is the $KN$ sigma
term.  This remains attractive for $K^- N$
interactions. The virtual pair term
removes\cite{BKR}, at zero temperature, about 60 \%
of the attraction (\ref{attraction}). At low temperature, the net effect is
therefore highly repulsive for $K^+ N$ interactions.

What happens as $T\rightarrow T_{\chi SR}$ is as follows. First of all,
part of the virtual pair repulsion gets ``boiled" off as discussed in
\cite{BKR}. More
importantly, if the Georgi vector limit is relevant, then
the vector mesons decouple with $g\rightarrow 0$, removing the repulsion
(\ref{repulsion}). As a consequence, the residual attraction from the
scalar exchange remains. The calculation of Koch\cite{koch} supports this
scenario. Since the vector coupling is absent,
both $K^+$ and $K^-$ will have a similar cool component.

\subsection*{Acknowledgments}
\indent

I am grateful for discussions with Gerry Brown, Youngman Kim and
Hyun Kyu Lee.

\end{document}